\documentclass[aoas,MSNbibl,nameyear,rotating,dvips]{arximspdf}
\usepackage{dcolumn}
\usepackage{graphicx}

\doi{10.1214/12-AOAS594} 
\volume{7}
\issue{1}
\pubyear{2013}
\firstpage{471}
\lastpage{494}

\makeatletter
\newcolumntype{d}[1]{D{.}{.}{#1}}
\newcommand{\rrvert}{\vert}
\newcommand{\mH}{\mathcal{H}}
\renewcommand{\P}{\mathrm{P}}
\newcommand{\IP}{\operatorname{IP}}
\newcommand{\C}{\mathrm{C}}
\renewcommand{\L}{\mathrm{L}}
\newcommand{\eqref}[1]{(\ref{#1})}
\newproclaim{alg}{Algorithm}
\makeatother

\begin{document}
\begin{frontmatter}

\title{Multiple testing of local maxima for detection of~peaks in ChIP-Seq data}
\runtitle{Multiple testing of local maxima in ChIP-Seq}

\begin{aug}
\author[A]{\fnms{Armin}~\snm{Schwartzman}\corref{}\thanksref{t1}\ead[label=e1]{armins@hsph.harvard.edu}},
\author[B]{\fnms{Andrew}~\snm{Jaffe}\thanksref{t2}\ead[label=e2]{andrew.jaffe@libd.org}},
\author[A]{\fnms{Yulia}~\snm{Gavrilov}\thanksref{t1}\ead[label=e3]{yuliagavrilov@gmail.com}}
\and
\author[A]{\fnms{Clifford~A.}~\snm{Meyer}\ead[label=e4]{cliff@jimmy.harvard.edu}}
\thankstext{t1}{Supported in part by the Claudia Adams Barr Program in
Cancer Research,
the William F. Milton Fund and NIH Grants P01-CA134294 and R01-CA157528.}
\thankstext{t2}{Supported in part by T32 GM074906 Predoctoral
Biostatistics Training in Genetics/Genomics.}
\runauthor{Schwartzman, Jaffe, Gavrilov and Meyer}
\affiliation{Harvard School of Public Health and Dana-Farber Cancer Institute,
Lieber~Institute for Brain Development and Johns Hopkins Bloomberg
School of Public Health, Harvard School of Public Health and Dana-Farber~Cancer Institute,
and Harvard School of Public Health and~Dana-Farber Cancer Institute}
\address[A]{A. Schwartzman\\
Y. Gavrilov\\
C.~A. Meyer\\
Department of Biostatistics\\
Harvard School of Public Health\\
and\\
Department of Biostatistics\\
\quad and Computational Biology\\
Dana-Farber Cancer Institute\\
450 Brookline Ave., CLS-11007\\
Boston, Massachusetts 02115\\
USA\\
\printead{e1}\\
\phantom{E-mail:\ }\printead*{e3}\\
\phantom{E-mail:\ }\printead*{e4}}
\address[B]{A. Jaffe\\
Lieber Institute for Brain Development \\
855 North Wolfe Street\\
Baltimore, Maryland 21205\\
USA\\
and\\
Department of Biostatistics\\
Johns Hopkins Bloomberg \\
\quad School of Public Health\\
615 North Wolfe Street\\
Baltimore, Maryland 21205\\
USA\\
\printead{e2}}
\end{aug}

\received{\smonth{8} \syear{2010}}
\revised{\smonth{8} \syear{2012}}

%
\begin{abstract}
A topological multiple testing approach to peak detection is proposed
for the problem of detecting transcription factor binding sites in
ChIP-Seq data. After kernel smoothing of the tag counts over the
genome, the presence of a peak is tested at each observed local
maximum, followed by multiple testing correction at the desired false
discovery rate level. Valid $p$-values for candidate peaks are computed
via Monte Carlo simulations of smoothed Poisson sequences, whose
background Poisson rates are obtained via linear regression from a
Control sample at two different scales. The proposed method identifies
nearby binding sites that other methods do not.
\end{abstract}

%

\end{frontmatter}

\section{Introduction}

The problem of detecting signal peaks in the presence of background
noise appears often in the analysis of high-throughput data. In
ChIP-Seq data, the problem of finding transcription factor binding
sites along the genome translates to a large-scale peak detection
problem with a one-dimensional spatial structure, where the number,
locations and heights of the peaks are unknown. Recently, \citet
{Schwartzman2011b} (hereafter SGA) introduced a topological multiple
testing approach to peak detection where, after kernel smoothing, the
presence of a signal is tested not at each spatial location but only at
the local maxima of the smoothed observed sequence. In this paper, we
show how that approach can be used to formalize the inference problem
of finding binding sites in ChIP-Seq data. To achieve this, we also
propose a new regression-based method for estimating the local
background binding rate from a Control sample.

\subsection{ChIP-Seq data}
\begin{figure}

\includegraphics{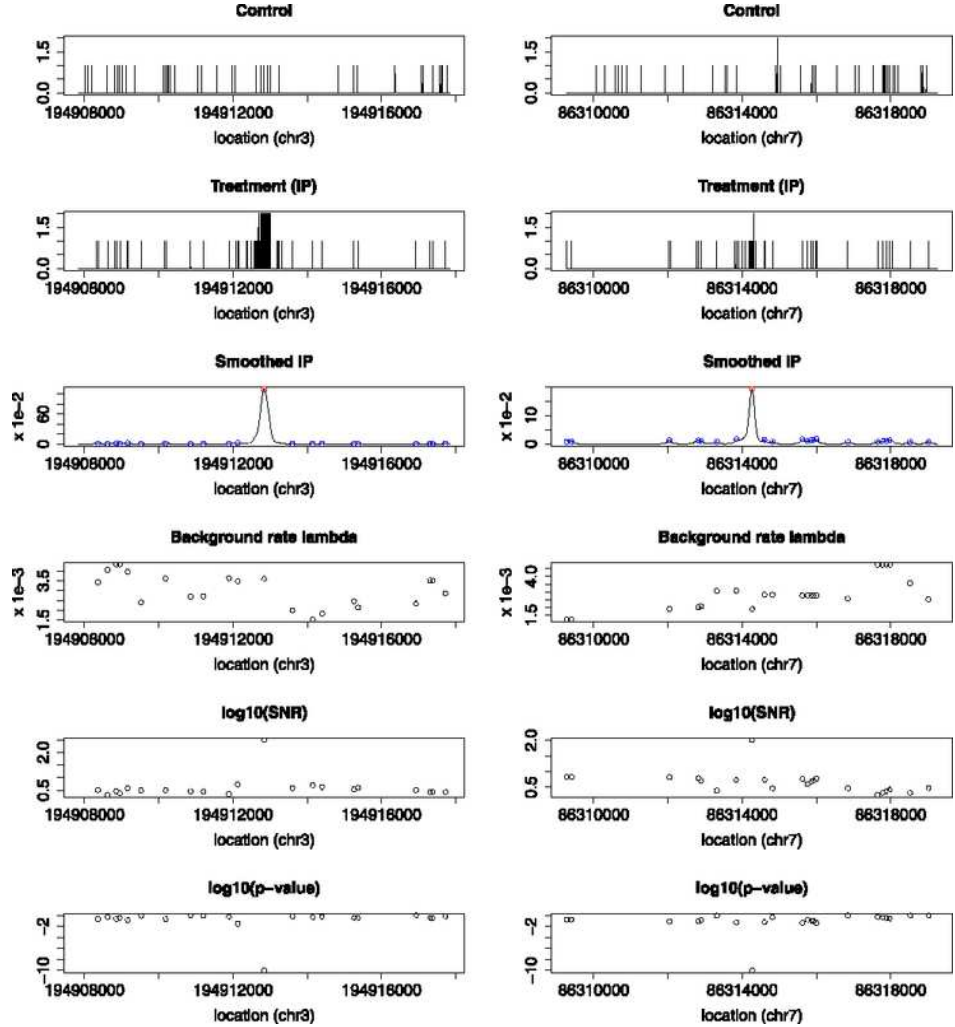}

\caption{A fragment of the Fox A1 aligned data featuring a few
representative peaks found by our method. Row 1: Control sample. Row 2:
IP sample, same fragment as the Control.
Row 3: Smoothed IP sample; significant peaks are indicated in red,
nonsignificant ones in blue. Row 4: Estimates
of the background Poisson rate $\lambda_0(t)$ at local maxima of the
smoothed IP sample. Row 5: Signal-to-noise
ratio (SNR), equal to peak height divided by background rate (log 10
scale). Row 6: $P$-values (log~10 scale).
Notice the difference in vertical scales between the left and right
panels.}\label{figFoxA1examples}\vspace*{-3pt}
\end{figure}

ChIP-Sequencing or ChIP-Seq is an experimental method that is often
used to map the locations of binding sites of transcription factors
along the genome in vivo [\citet{Barski2009,Park2009}]. Transcription
factors control the transcription of genetic information from DNA to
mRNA in living cells, and abnormalities in this process are often
associated with cancer. Given a particular transcription factor of
interest, ChIP-Seq combines chromatin immunoprecipitation (ChIP) with
massively parallel DNA sequencing, allowing enrichment of the DNA
segments bound by the transcription factor and mapping of their
locations along the genome. The result is a long list of sequenced
forward and reverse tags, also called reads, each associated with a
specific genomic address. After alignment of these tags, the data
consists of a sequence of tag counts along the genome, with a tendency
to a higher concentration of tags near the transcription factor binding
sites. An example of a data fragment is shown in Rows 1 and~2 of Figure
\ref{figFoxA1examples}. (Note that not all ChIP-Seq data follow this
pattern, e.g., histone modification data.)

The goal of the analysis is to identify the true binding sites. This
translates to finding genomic locations where the binding rate is
higher than it would be if the transcription factor were not present.
To this end, \citet{Johnson2007} suggested sequencing a Control input
sample to provide an experimental assessment of the background tag
distribution, helping reduce false positives. The cost currently
associated with this technology often does not allow more than a single
ChIP-Seq sample, also called an IP sample, and a single Control sample.
To illustrate the usefulness of the Control, Rows 1 and 2 of Figure \ref
{figFoxA1examples} show a short fragment of the raw data after
alignment in the Control and IP samples, respectively, for the same
positions in the genome. The interesting peaks are marked by red
circles in Row 3, corresponding to sites with high binding rate in the
IP sample but lower rate in the Control. Other candidate peaks, marked
in blue, do not have a significantly higher binding rate in the IP
sample than in the Control.

%

As an additional condition, it is necessary that a site has a high
binding rate in absolute terms to avoid spurious high fold enrichments
due to high variability at low coverage (e.g., 3-fold enrichment
resulting from 3 reads in treatment vs. 1 read in control).

\subsection{Testing of local maxima}
The search for binding sites may be set up as a large-scale multiple
testing problem where, at each genomic location, a~test is performed
for whether the binding rate is higher than the background. Testing at\vadjust{\goodbreak}
each genomic location is statistically inefficient because it requires
a multiple testing correction for a very large number of tests over the
entire length of the genome. In ChIP-Seq, the binding rate at a true
binding site has a unimodal peak shape that spreads into neighboring
locations, caused by the variability in the start and end points of the
sequenced segments. Thus, as argued by SGA, it is enough to test for
high binding rates only at locations that resemble peaks, that is,
local maxima of the smoothed data. In this sense, the local maxima
serve as topological representatives of the candidate binding sites.

Peak detection on the aligned data is carried out using the Smooth and
Test Local Maxima (STEM) algorithm of SGA. It consists of the following:
\begin{longlist}[(1)]
\item[(1)] kernel smoothing;
\item[(2)] finding the local maxima as candidate peaks;
\item[(3)] computing $p$-values for the heights of the observed local maxima; and
\item[(4)] applying a multiple testing procedure to the obtained $p$-values.
\end{longlist}
For Step 1, following the ``matched filter principle'' recommended by
SGA, we use a symmetric unimodal kernel that roughly matches the shape
of the peaks to be detected. This shape corresponds to the spatial
spread of tag locations around a true binding site and is assumed to be
the same for all binding sites, up to an amplitude scaling factor
dictated by the physics and chemistry of the experimental protocol.
This shape, up to an amplitude scaling factor, is estimated from the
data during the alignment process. In Step~2, local maxima are defined
as smoothed counts that are higher than their neighbors after
correcting for ties. In Step 3, $p$-values test the hypothesis that the
local binding rate is less or equal to the local background rate or a
minimally interesting binding rate. The required distribution of the
heights of local maxima is computed via Monte Carlo simulations.
Finally, Step 4 is carried out using the Benjamini--Hochberg (BH)
procedure [\citet{Benjamini1995}], although, in general, other multiple
testing algorithms may be used instead.

The STEM algorithm is promising for ChIP-Seq data because it was shown
in SGA to provide asymptotic error control and power consistency under
similar modeling assumptions. Like in ChIP-Seq data, SGA assumed that
the signal peaks are unimodal with finite support and that the search
occurs over a long observed sequence. Further assuming additive
Gaussian stationary ergodic noise, SGA proved that the BH procedure
controls the false discovery rate (FDR) of detected peaks, defined as
the expected ratio of falsely detected peaks among detected peaks,
where a detected peak is considered true (false) if it occurs inside
(outside) the support of any true peak. In SGA, the control is
asymptotic as both the search space and the signal strength increase,
where the former may grow exponentially faster than the latter, and the
detection power tends to one under the same asymptotic conditions. In
ChIP-Seq data, the definitions of true and false detected peaks apply
within the spatial extent of the true peak shape, which is estimated
here during the alignment process.

\subsection{Estimation of the background rate and Monte Carlo
calculation of $p$-values}
ChIP-Seq data differs from the modeling assumptions of SGA in that
ChIP-Seq data\vadjust{\goodbreak} consists of a long sequence of positive integer counts,
often assumed to follow a Poisson distribution [\citet{Mikkelsen2007}].
Moreover, the process generating the background noise counts is not
globally stationary [\citet{Johnson2007}]. To make inference possible,
we assume the background Poisson rate to vary over the genome but not
too fast so that it is approximately constant in the immediate vicinity
of any candidate peak. The background Poisson rate at any given
location is estimated as a linear function of the local Control counts
at two different spatial scales, 1~kilo base-pairs (kb) and 10~kb. The
linear coefficients are estimated from the data by multiple regression,
automatically solving the normalization problem of having different
sequencing depths between the IP and Control samples.

Finally, as required by Step 3 of the STEM algorithm above, for an
observed local maximum of the smoothed ChIP-Seq data at a given
location, its $p$-value is computed via Monte Carlo simulation using the
background Poisson parameter estimated for that location. Note that the
STEM algorithm requires an estimate of the background, but does not
depend on how that estimate was obtained. Here we propose a regression
method, but that method could be changed without changing the basic
operation of the STEM algorithm.

\subsection{Other methods}
Several ChIP-Seq data analysis methods have been proposed in the
literature; cf. MACS [\citet{MACS2008}], cisGenome [\citet
{cisGenome2008}], QuEST [\citet{QuEST2008}] and FindPeaks [\citet
{Fejes2008}]. While these methods also view the problem of detecting
binding sites as a peak detection problem, use statistical models and
estimate error rates, most of them do not formally state the
statistical inference problem. Exceptions are PICS [\citet{PICS2010}]
and BayesPeak [\citet{BayesPeak2009}], which are both Bayesian
approaches, whereas we adopt a frequentist point of view. QuEST [\citet
{QuEST2008}] also finds local maxima as candidate peaks but uses a
narrow Gaussian kernel rather than a matched filter and estimates the
FDR by comparing the number of peaks called in the IP and Control
sequences rather than estimating the background and formally testing
using $p$-values. T-PIC [\citet{T-PIC2011}] also takes a topological
approach, but rather than heights of local maxima it measures the depth
of trees built from excursion regions of the coverage function of the
data, so our method is simpler.

Here we attempt to frame the ChIP-Seq analysis problem as a formal
inference problem in multiple testing relying on the error control
properties proven in SGA and using a new regression method to estimate
the background binding rate. As a reference, we compare the results of
our analysis to those of MACS, cisGenome and QuEST on two different
data sets. By focusing on detecting peaks rather than regions and using
a matched filter, our approach has the ability to distinguish nearby
binding sites that MACS and ciSGenome do not, and in a less fragmented
fashion than QuEST.

\subsection{Data sets}
We demonstrate our approach on two different ChIP-Seq data sets. In the
first, ChIP-Seq targeting the transcription factor FoxA1 was performed
on the breast cancer cell line MCF-7 [\citet{MACS2008}]. This data set
includes a ChIP-Seq sample (hereafter IP), in which the FoxA1 antibody
was used, and a Control input sample, in which the procedure was
repeated without the antibody. Sequencing covered the entire genome,
producing about 3.9 million tags in the IP sample and about 5.2 million
tags in the Control sample. The second data set concerns the
growth-associated binding protein (GABP) [\citet{QuEST2008}]. This
larger data set consists of an IP sample with about 7.8 million tags
and a Control sample with about 17.4 million tags. The methods in this
paper were developed on the FoxA1 data set and later applied to the
GABP data set as an independent testbed. In both data sets, the goal of
the analysis is to detect genomic loci in the IP sample that have a
significantly high number of tags both in absolute terms and relative
to the Control sample.

It should be noted that the goal of this paper is not to propose a new
peak finding tool, but rather to show how a topological inference
approach can be used to provide formal statistical inference in
ChIP-Seq data, with the view that its basic principles can be
generalized to other genomic search problems [\citet{Jaffe2012}].
The methods in this paper were implemented in \texttt{R}.

\section{Peak detection for ChIP-Seq data}
\label{secmain}

\subsection{Alignment and estimation of the peak shape}
\label{secalignment}

Before statistical analysis, we follow the approach in MACS of first
aligning the forward and reverse tags, after which tags can be treated
indistinctively. The alignment process, described in the \hyperref[app]{Appendix}, also
allows us to estimate the amount by which tags need to be shifted and
the shape of the spatial spread of the shifted tag counts around a peak.

\begin{figure}

\includegraphics{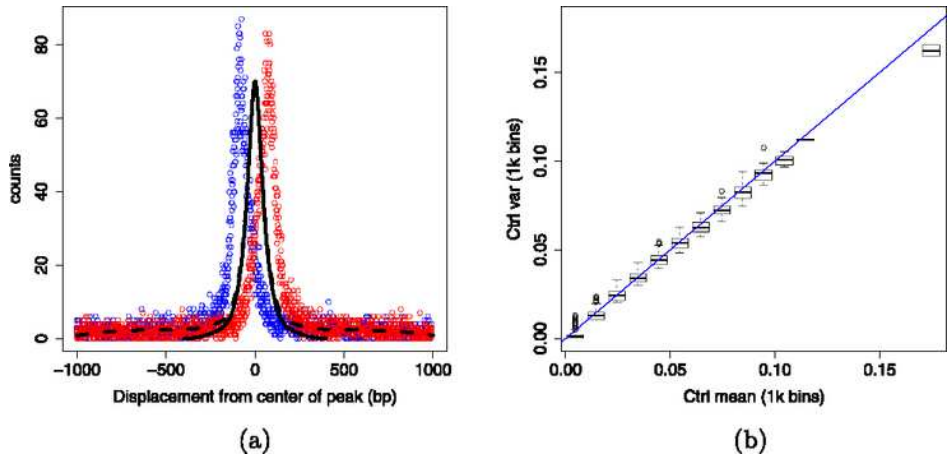}

\caption{\textup{(a)} Estimated distribution of tag counts in the
forward strand (red) and in the reverse strand (blue) of the FoxA1 data set
(chromosome 1). Aligning the distributions and averaging the counts
results in
the joint count distribution and peak shape (black dashed). The peak shape
is multiplied by a quartic biweight function (black solid). \textup{(b)} Sample
mean vs.
sample variance of the aligned Control sequence in bins of size 1~Kb.
The blue line has slope 1.}\label{figpeak-shape}
\end{figure}

For illustration, Figure~\ref{figpeak-shape}(a) shows the spatial
distributions of the forward and reverse tags in the IP sample of the
FoxA1 data set before alignment, obtained from 1000 strong and easily
detectable peaks in chromosome 1. These distributions are displaced
with respect to one another. The optimal shift found in this case was
62 base pairs (bp), almost the same as the estimated shift of 63 found
by MACS for the same data. Shifting the distributions by this amount
produces the black-dashed overlap distribution shape.

As a further refinement, Figure~\ref{figpeak-shape}(a) shows that the
binding rate is approximately constant beyond about 400~bp away from
the peak center, and hence should not be included as part of the peak.
As a correction, the support of the estimated peak shape was reduced by
multiplying the black-dashed shape by a quartic biweight function of
size $W = 801$, producing the estimate in solid black. This peak shape,
normalized to unit sum, is used as a smoothing kernel in the STEM
algorithm for peak detection. The quartic biweight function has the
effect of providing the kernel with continuous derivatives at the
edges, a desirable property to avoid spurious local maxima at that step
of the algorithm.

%

\subsection{The Poisson model and the STEM algorithm}
After alignment, the data consists of a table of genomic locations,
each with an associated tag count. The remaining genomic locations are
assumed to have a count of zero. Since the data is given as positive
integer counts, it is reasonable to model them as Poisson variables
[\citet{Mikkelsen2007}]. Specifically, we assume that the IP and
Control counts $\IP(t)$ and $\C(t)$ at locations $t$ are independent
Poisson sequences
%
\begin{equation}
\label{eqPoisson-model} \IP(t)\sim \operatorname{Po} \bigl[\lambda_{\IP}(t) \bigr],\qquad
\C(t)\sim \operatorname{Po} \bigl[\lambda_{\C}(t) \bigr],\qquad t \in\mathbb{Z},
\end{equation}
where $\lambda_{\IP}(t) \ge0$ and $\lambda_{\C}(t) \ge0$ denote the
mean rates at location $t$, which may vary over $t$. The values of the
processes $\IP(t)$ and $\C(t)$ are assumed independent over $t$ given
$\lambda_{\IP}(t)$ and $\lambda_{\C}(t)$.

As model validation, Figure~\ref{figpeak-shape}(b) shows a graph of the
sample mean vs. sample variance of the aligned Control sequence in the
FoxA1 data set, computed in bins of size 1 kbp. The two quantities are
nearly proportional with a proportionality constant of 1, as expected
from the Poisson model~\eqref{eqPoisson-model}. The IP sample exhibits
a similar pattern (not shown).

Regions of high binding frequency are represented by peaks in the mean
Poisson rates. The goal is to find regions where $\lambda_{\IP}(t)$ is
higher than the local background rate $\lambda_0(t)$, but also higher
than a minimal constant binding rate $\lambda_{\L}$. The lower bound
$\lambda_{\L}$ avoids detecting spurious weak peaks in the presence of
an even weaker local background. Here $\lambda_{\L}$ is set to the
global average rate, equal to the total number of aligned tags in the
IP sequence divided by the total length of the genome. Taking the
latter as $3.018 \times10^9$ [\citet{Sakharkar2004}], the global
average rate for the FoxA1 data set is $\lambda_{\L} = (3.57\times
10^6)/(3.018\times10^9) = 0.00118$.

At every $t$, the above comparison translates to testing whether
$\lambda_{\IP}(t) \le\lambda_0(t)$ and $\lambda_{\IP}(t) \le\lambda_{\L}$, that is, $\lambda_{\IP}(t) \le\max\{\lambda_0(t), \lambda_{\L
}\}$. To gain efficiency, rather than testing at every single location
$t$, tests are performed at only local maxima of the smoothed IP
sequence. This is carried out formally using the following adaptation
of the STEM algorithm from SGA.

\begin{alg}[(STEM algorithm)]
\label{algSTEM}
\begin{longlist}[(1)]
\item[(1)] Let $w(t)$ be a unimodal kernel of length $W$. Apply kernel
smoothing to the IP sequence to produce the smoothed sequence
%
\begin{equation}
\label{eqIP-gamma} \widetilde{\IP}(t) = w(t) * \IP(t) = \frac{1}{W}\sum
_{s=-(W-1)/2}^{(W+1)/2} w(s) \IP(t-s).
\end{equation}
\item[(2)] Find all local maxima of $\widetilde{\IP}(t)$ as candidate peaks.
Let $\tilde{T}$ denote the set of locations of those local maxima.
\item[(3)] For each local maximum $t \in\tilde{T}$, compute a $p$-value
$p(t)$ for testing the null hypothesis
%
\begin{equation}
\label{eqH0} \mH_0(t)\dvtx \lambda_{\IP}(t) \le
\lambda_0^+(t) \quad\mbox{vs.}\quad \lambda_{\IP}(t) >
\lambda_0^+(t)
\end{equation}
in a neighborhood of $t$, where $\lambda_0^+(t) = \max\{\lambda_0(t),
\lambda_{\L}\}$.
\item[(4)] Let $\tilde{m}$ be the number of local maxima. Apply a multiple
testing procedure on the set of $p$-values and declare significant all
peaks whose $p$-values are smaller than the threshold.
\end{longlist}
\end{alg}

Details on each of the steps are given in the following sections.

\subsection{Smoothing and local maxima}
\label{secsmoothing}

According to SGA, the best smoothing kernel for the purposes of peak
detection is that which maximizes the signal-to-noise ratio (SNR) after
convolving the peak shape, assumed to underly the signal peaks in the
data, with the smoothing kernel. This is achieved by choosing the
smoothing kernel to be equal to the peak shape itself (up to a scaling
factor), a principle long known in signal processing as ``matched
filter theorem'' [\citet{North1943,Turin1960,Pratt1991,Simon1995}].
Note that this is not the same as the optimal kernel in nonparametric
regression [\citet{Wasserman2006}].\vadjust{\goodbreak}

In ChIP-Seq data, binding rate peaks corresponding to different binding
sites for the same transcription factor are assumed to have the same
shape in terms of spatial spread, but may have different heights. The
common peak shape is estimated in the alignment process (solid curve in
Figure~\ref{figpeak-shape}). It is unimodal, constrained to be
symmetric, and has heavier tails than the Gaussian density. In Step 1
of the STEM algorithm (Algorithm~\ref{algSTEM}), smoothing was carried
out setting $w(t)$ equal to the solid curve in Figure \ref{figpeak-shape}, normalized to have unit sum, with $W = 801$.

Rows 2 and 3 in Figure~\ref{figFoxA1examples} compare the raw and
smoothed IP data. The smoothed data is high at locations where the
density of tag counts is high. Notice that kernel smoothing produces
positive counts locations where the unsmoothed IP data may have no counts.

In Step 2 of the STEM algorithm (Algorithm~\ref{algSTEM}), local
maxima of the smoothed sequence $\widetilde{\IP}(t)$ are defined as
values $\widetilde{\IP}(t)$ that are greater than their immediate
neighbors $\widetilde{\IP}(t-1)$ and $\widetilde{\IP}(t+1)$. If the
maximum is tied between neighboring values, then the peak location is
assigned the lower genomic address. A useful property of the kernel
that avoids producing spurious local maxima is to have continuous
derivatives. This was ensured by multiplication of the estimated peak
shape by a quartic biweight function, as described in Section \ref
{secalignment} above.

Restricting the analysis to local maxima reduces the amount of data to
process further. In the aligned IP sample of the FoxA1 data set, the
number of local maxima found was about 2.7 million, down from about 3.9
million original mapped tags.

\subsection{Estimation of the local background rate}
\label{secparams}

Computation of $p$-values in Step~3 of the STEM algorithm (Algorithm \ref
{algSTEM}) requires knowledge of the background Poisson rate $\lambda_0(t)$
under the null hypothesis. Estimation of $\lambda_0(t)$ is
difficult because it varies with $t$ in an unknown fashion [\citet
{Johnson2007}]. Here we propose a simple method to estimate the
background rate from the local Control data, as follows.

Since the Control sample is intended to represent the background
process in the IP sample, it is reasonable to assume that the local
background rate $\lambda_0(t)$ in the IP sample is proportional to the
corresponding local background rate $\lambda_{\C}(t)$ in the Control
sample, reflecting the ratio in sequencing depth of the background
between the two samples. In the FoxA1 data, the IP sample has about 3.9
million tags, while the Control sample has about 5.2 million counts.

The local Control rate $\lambda_{\C}(t)$, in turn, may be estimated as
the average tag count in the Control sample within a certain window
centered at $t$, as in kernel-based nonparametric regression methods
[\citet{Wasserman2006}]. The window size establishes a
bias-vs.-variance trade-off in the estimation. While the background
rate may change fast,\vadjust{\goodbreak} 1~Kb is about the smallest window size that
allows comparison of peaks, usually of size a few hundred~bp, against
the background. Because counts are often sparse, to add stability to
the parameter estimates, we consider the local rate to be also linearly
related to the corresponding rate in the Control within a window of
size 10~Kb centered at $t$.

\begin{figure}

\includegraphics{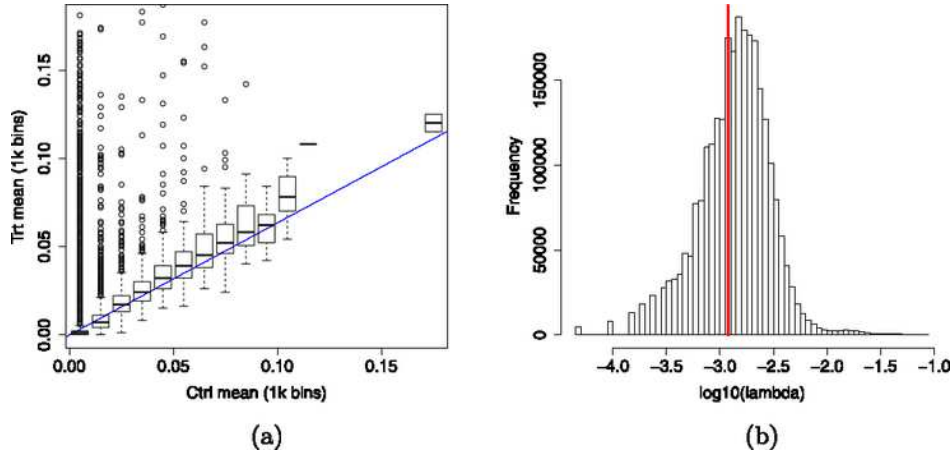}

\caption{Marginal distributions of the 1~Kb bin
averages in the
IP sample of the FoxA1 data set as a function of: \textup{(a)} The 1~Kb bin
averages in the Control sample; \textup{(b)} Histogram of estimated values of
$\hat{\lambda}_0$ in the FoxA1 data set. As a reference, the global
average rate is 0.00118 (red).}\label{figregression}\vspace*{-3pt}
\end{figure}

To illustrate these relationships, Figure~\ref{figregression}(a) shows a
graph of the 1~Kb bin averages in the Control sample of the FoxA1 data
set against the 1~Kb bin averages in the IP sample. While there is a
lot of variability, the main trend is seen to be linear, captured in
the figure by a marginal linear fit. The outliers in the upper left
corner correspond vaguely to the peaks sought. However, their
relatively small number introduces little bias in the regression. A
similar trend is observed when plotting the 1~Kb bin averages in the IP
sample as a function of the 10~Kb bin averages in the Control sample
(not shown).

%

Summarizing, $\lambda_0(t)$ is estimated from local windows of sizes 1
Kb and 10~Kb centered at $t$ via
%
\begin{equation}
\label{eqlambda0} \hat{\lambda}_0(t) = a_1 \hat{
\lambda}_{\C,1k}(t) + a_2 \hat{\lambda }_{\C,10k}(t),
\end{equation}
where $\hat{\lambda}_{\C,1k}(t)$ and $\hat{\lambda}_{\C,10k}(t)$ are
the Control averages in windows of size 1~Kb and 10~Kb centered at $t$,
and $a_1$, $a_2$ are global parameters. Note that the combination of
the 1~Kb and 10~Kb windows plays the role of trapezoidal kernel whose
shape is optimally determined by the data-determined coefficients $a_1$
and $a_2$. To estimate $a_1$ and $a_2$, we set up a global linear
regression as in \eqref{eqlambda0}, except that the predictors and the
response are replaced by the 1~Kb and 10~Kb bin averages, as in
Figure~\ref{figregression}(a).\vadjust{\goodbreak}

Applying this regression in the FoxA1 data set gave estimates $\hat
{a}_1 = 0.307 \pm0.001$ and $\hat{a}_2 = 0.482 \pm0.001$, giving more
weight to the 10~Kb window than the 1~kb window. The coefficients
automatically account for sequencing depth: if~the binding rate in the
Control were constant, then the background estimate for the IP would be
approximately equal to the Control rate multiplied by the sum of the
two window coefficients, equal to 0.789. This factor is slightly
smaller than the overall ratio between the total number of aligned
counts in the IP sample and in the Control sample, equal to 0.805. The
extra counts in the IP sequence are precisely the signal we wish to detect.

The multi-window model makes the estimate adaptive to the local
variability in the background rate. As an example, Row 4 of Figure \ref
{figFoxA1examples} shows the local estimates $\hat{\lambda}_0(t)$,
roughly following the tag pattern observed in Row 1. Row 5 shows the
SNR, defined as the ratio between the peak height $\widetilde{\IP}(t)$
and the estimated background rate $\hat{\lambda}_0(t)$. Figure \ref
{figregression}(a) shows the distribution of the estimated values of
$\hat{\lambda}_0(t)$ over the entire genome for the FoxA1 data
set.\looseness=-1

\subsection{Computing $p$-values}

In Step 3 of Algorithm~\ref{algSTEM}, the $p$-value $p(t)$ of an
observed local maximum of the smoothed sequence $\widetilde{\IP}(t)$ at
a location $t$ is defined as the probability to obtain the observed
height of the local maximum or higher under the least favorable null
hypothesis $\lambda_{\IP}(t) = \lambda_0^+(t)$ in \eqref{eqH0}. The
null hypothesis need only be assumed in a local neighborhood of each
candidate peak because $\widetilde{\IP}(t)$ depends only on the data
within a local neighborhood, as dictated by the smoothing kernel
$w(t)$. In this section we assume that $\lambda_0(t)$ and $\lambda_{\L
}$ are known, having been estimated according to the methods described
in Section~\ref{secparams} above.

In SGA, the background noise process was assumed stationary. In
ChIP-Seq data, in contrast, the background rate $\lambda_0(t)$ is not
constant. However, if the background process is locally stationary,
then the background process in the neighborhood of a given location $t
= \tilde{t}$ may be assumed to have similar statistical properties in
that neighborhood as a stationary sequence with constant background
rate $\lambda\equiv\lambda_0(\tilde{t})$. In particular, the height
of a local maximum of the smoothed sequences at $\tilde{t}$ would have
approximately the same distribution in both cases.

Specifically, suppose $X(t; \lambda)$ is a sequence of i.i.d. Poisson
random variables with constant mean rate $\lambda$. Smoothing of $X(t;
\lambda)$ with the kernel $w(t)$ as in Step 1 of Algorithm \ref
{algSTEM} produces the smoothed sequence
%
\begin{equation}
\label{eqXtilde} \tilde{X}(t; \lambda) = w(t) * X(t; \lambda) =
\frac{1}{W}\sum_{s=-(W-1)/2}^{(W+1)/2} X(t-s;
\lambda).
\end{equation}
The height of a local maximum of the stationary sequence $\tilde{X}(t;
\lambda)$ has the survival function
%
\begin{equation}
\label{eqdistr-max} F(u; \lambda) = \P \bigl[  \tilde{X}(t; \lambda) \ge
u  \rrvert   t\mbox{ is a local maximum}, \lambda \bigr].
\end{equation}
Then, the null distribution of the height of a local maximum of
$\widetilde{\IP}(t)$ at $t$ may be approximated by the distribution
$F(u; \lambda)$ \eqref{eqdistr-max} corresponding to the constant rate
$\lambda\equiv\lambda_0(t)$. Finally, given the observed height
$\widetilde{\IP}(t)$ at $t$, its $p$-value under the null hypothesis
\eqref{eqH0} is defined as
%
\begin{equation}
\label{eqpv} p(t) = F \bigl(\widetilde{\IP}(t); \lambda_0^+(t)
\bigr).
\end{equation}

The distribution \eqref{eqdistr-max} is difficult to compute
analytically. Instead, we resort to Monte Carlo simulations, where for
each given value of $\lambda$, a long sequence $X(t; \lambda)$ of
i.i.d. Poisson variables is generated, smoothed using the kernel
$w(t)$, and its local maxima found. The distribution \eqref
{eqdistr-max} is then estimated empirically from the obtained heights
of the local maxima of the smoothed simulated sequence $\tilde{X}(t;
\lambda)$ \eqref{eqXtilde}.

To reduce computations, rather than performing a new simulation for
each new background rate $\lambda_0^+(t)$, a table of survival
functions \eqref{eqdistr-max} is prepared in advance for a set of
values of $u$ and $\lambda$ that covers the range of possible values to
be found in the data. Then, to evaluate $\hat{F}(u; \lambda)$ in \eqref
{eqpv} for any particular pair of values of $\widetilde{\IP}(t)$ and
$\lambda_0(t)$, bilinear interpolation is used between the closest grid points.

In the FoxA1 data set, the smallest and largest values of $\hat{\lambda
}_0(t)$ found were $1.67\times10^{-5}$ and $7.40\times10^{-2}$,
respectively, giving values of $\hat{\lambda}_0^+(t)$ in the range
0.00118 to 0.0740. Taking a safety margin of 25\%,
we performed the Monte Carlo simulation described above for 300 values
of $\lambda$ equally spaced on a logarithmic scale between 0.00089 and
0.0925. The length of the simulated Poisson sequences was set to be as
long as needed to obtain at least 100 nonzero counts, but not smaller
than $1\times10^5$. In order to reduce the variability from the
simulation, the table of survival functions $\hat{F}(u,\lambda)$ was
smoothed over $\lambda$ for each fixed $u$ via linear regression using
5 B-spline basis functions. The 25\%
safety margins ensured that none of the values of $\lambda$ actually
needed were near the edges of the table for the purposes of spline
smoothing.\looseness=-1

\begin{figure}

\includegraphics{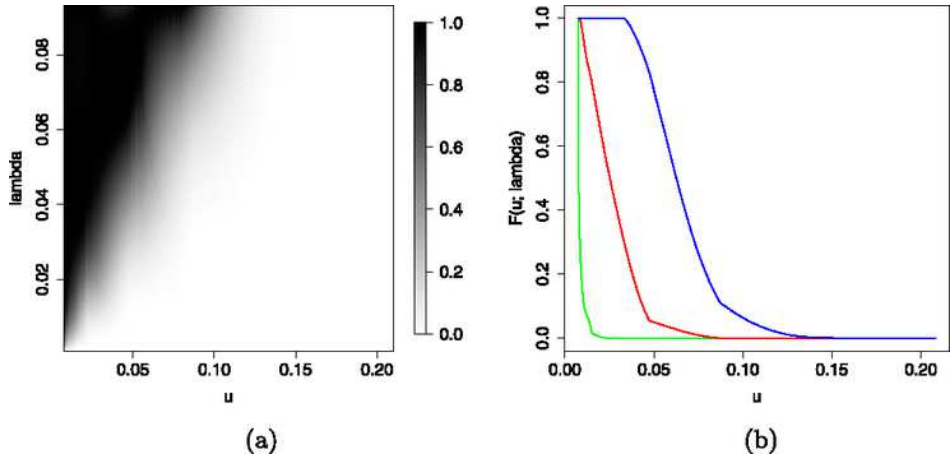}

\caption{Survival functions $\hat{F}(u; \lambda)$
of the height $u$ of local maxima, approximated by Monte Carlo simulation.
\textup{(a)} Viewed in gray scale as a function of the background rate $\lambda$.
\textup{(b)} Specific survival functions for $\lambda= \lambda_L = 0.0018$ (green),
$\lambda= 0.020$ (red), and $\lambda= 0.056$ (blue).}\label{figlambda-distr}
\end{figure}

Figure~\ref{figlambda-distr} shows the obtained function $\hat{F}(u;
\lambda)$ \eqref{eqdistr-max}, given as a table of size 300 values of
$\lambda$ by 200 values of $u$ and for a few particular values of
$\lambda$. As an example, Row 6 of Figure~\ref{figFoxA1examples}
shows the calculated $p$-values $p(t)$ in the corresponding data
segments. Because of numerical precision in the Monte Carlo
simulations, very low $p$-values could not be distinguished from zero,
and in the figures they are drawn as if they were equal to $10^{-10}$.

%

Notice in Figure~\ref{figlambda-distr} that the smallest value of $u$
is 0.0076, which corresponds to the height of a local maximum obtained
from a single tag. Any isolated tag (farther than 1~kb from any other
tag) constitutes the smallest possible local maximum and thus gets a
$p$-value of 1 regardless of the estimated background rate. In this
sense, using the global average rate $\lambda_L = 0.00118$,
corresponding to about 1 tag per 1~Kb, as an absolute reference, is not
restrictive. However, the results are sensitive to the choice of
$\lambda_L$ in the sense that,\vadjust{\goodbreak} if $\lambda_L$ is larger than the
estimated background rate $\lambda_0(t)$ at any location~$t$, then
$\lambda_0^+(t) = \lambda_L$ is used as the rate for the null
hypothesis rather than the estimated local background rate $\lambda_0$.
This can affect the significance of stronger peaks and it is therefore
preferable to choose a value of $\lambda_L$ that is not large, as it is
done here.

\subsection{Multiple testing}

Following SGA, we applied the BH procedure on the sequence of $\tilde
{m}=2\mbox{,}643\mbox{,}095$ $p$-values from the FoxA1 data set, each corresponding to a
local maximum of the smoothed sequence $\widetilde{\IP}(t)$. Of these
local maxima, 21,986 were declared significant at an FDR level of 0.01.
Their associated addresses $t$ are effectively point estimates of the
locations of the binding sites they represent.

%
\begin{figure}

\includegraphics{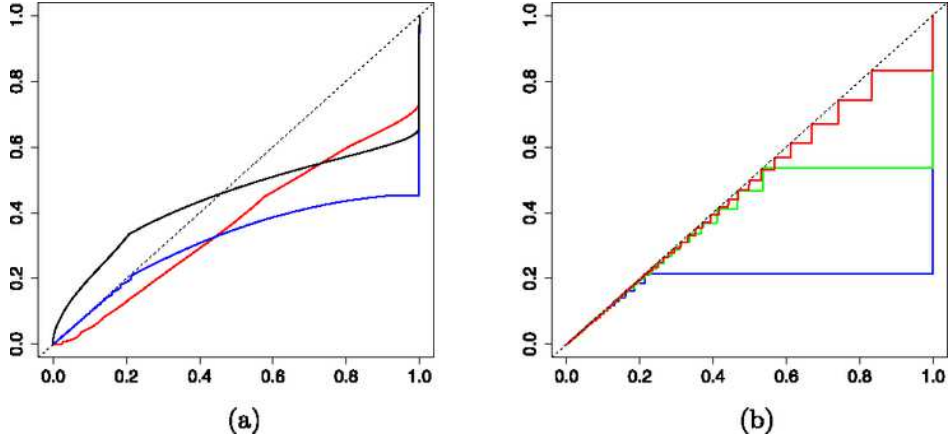}

\caption{Marginal distribution of $p$-values in the FoxA1 data set:
\textup{(a)} observed (black) and estimated under the global null hypothesis
empirically (red)
and theoretically (blue); \textup{(b)} specific null distributions for $\lambda
=0.0012$ (blue),
$\lambda= \lambda_{\L} = 0.0018$ (green), and $\lambda= 0.0030$
(red).}\label{figpv}
\end{figure}

As an example, in Row 3 of Figure~\ref{figFoxA1examples}, the
significant local maxima are indicated by red circles. As final
results, the detected peaks were ranked according to their $p$-values. Of
the 21,986 significant peaks, the top 7284 had $p$-values that could not
be distinguished from 0 because of the numerical accuracy of our Monte
Carlo simulations. These peaks were ranked according to their SNR.

To assess the validity of the procedure, Figure~\ref{figpv}(a) compares
the observed marginal distribution of $p$-values to the expected marginal
distribution under the complete null hypothesis in the FoxA1 data set.
The observed marginal distribution of $p$-values [shown in black in
Figure~\ref{figpv}(a)] is given by the empirical distribution
%
\begin{equation}
\label{eqobs-distr} \hat{G}(p) = \frac{1}{\tilde{m}}\sum
_{t \in\tilde{T}}\mathbf{1} \bigl[p(t) \le p \bigr],\qquad 0\le p\le1,\vadjust{\goodbreak}
\end{equation}
where $\tilde{T}$ is the set of $\tilde{m}$ locations of the local
maxima of $\widetilde{\IP}(t)$, with $p$-values given by \eqref{eqpv}.
The marginal distribution under the complete null hypothesis is
estimated in two different ways, one purely empirical and one more theoretical.

The empirical estimate [shown in red in Figure~\ref{figpv}(a)] was
obtained by running the entire analysis on the Control sample as if it
were the IP, that is, searching for peaks in the Control sample using
the same Control sample for estimating the background. The obtained
null distribution of $p$-values lies below the diagonal as required for
validity, and it exhibits a high frequency of $p$-values equal to 1,
corresponding to peaks with only one tag in them.

The theoretical estimate [shown in blue in Figure~\ref{figpv}(a)] was
obtained as follows. Recall that for a smoothed stationary Poisson
sequence $\tilde{X}(t;\lambda)$ with constant rate~$\lambda$, the
distribution of the height of a local maximum at $t \in\tilde{T}$ is
given by \eqref{eqdistr-max}. Analogous to \eqref{eqpv}, define the
corresponding null $p$-value as $p_0(t) = F (\tilde{X}(t); \lambda
)$ for $t\in\tilde{T}$. Its distribution $G_0(p; \lambda) =
P(p_0(t) \le p)$ for any $t$ is given by
%
\begin{equation}
\label{eqnull-distr} G_0(p; \lambda) = \cases{ 1, &\quad $F
(u_1; \lambda ) \le p$, \vspace*{2pt}
\cr
F (u_k; \lambda
), & \quad$F (u_k; \lambda ) \le p < F (u_{k-1}; \lambda ),
k=2,3,\ldots,$ }
\end{equation}
where $u_k$, $k=1,2,\ldots,$ are the discrete values taken by the
smoothed process $\tilde{X}(t; \lambda)$ at the local maxima. Note that
$G_0(p; \lambda)$ is independent of $t$ for $t\in\tilde{T}$ because of
stationarity. In the ChIP-Seq problem, we approximate the null
distribution of the $p$-value at $t \in\tilde{T}$ by the null
distribution $G_0(p; \hat{\lambda}_0^+(t))$ corresponding to a
stationary process with constant rate $\lambda= \hat{\lambda}_0(t)$,
which depends on $t$ only through the value of $\lambda$. Since each of
the observed $p$-values in \eqref{eqobs-distr} corresponds to a
different background rate $\hat{\lambda}_0(t)$, the estimated marginal
distribution under the global null hypothesis is given by the mixture
distribution
\begin{equation}
\label{eqnull-distr-global} \hat{G}_0(p) = \frac{1}{\tilde{m}}\sum
_{t \in\tilde{T}} G_0 \bigl(p; \hat{
\lambda}^+_0(t) \bigr),\qquad 0\le p\le1.
\end{equation}

Referring back to Figure~\ref{figpv}(a), the observed distribution is
always above the null distribution, and the large derivative at zero
indicates the presence of a strong signal, which explains the large
number of significant peaks found. Note that the null distribution is
not uniform but stochastically larger. To better understand the mixture
\eqref{eqnull-distr-global}, Figure~\ref{figpv}(b) shows three examples
of the individual null distributions \eqref{eqnull-distr}. All are
discrete and stochastically larger than the continuous uniform
distribution. For small $\lambda$, the most common $p$-value is 1, as
most local maxima take the smallest possible value $u_1$, equal to the
mode of the kernel $w(t)$, obtained when there is an isolated count of~1 in a neighborhood of zeros. This explains the large jump at~1 in
panel (a). As $\lambda$ gets larger, the distribution becomes closer to
the continuous uniform distribution.

\section{Comparison to other methods}
\label{seccomparison}

As a reference, we compared our meth\-od to MACS, cisGenome and QuEST on
both the FoxA1 and GABP data sets. While the FoxA1 data set was used in
the development of MACS and our method, the GABP data set was not used
in the development of any of the three methods, providing an
independent test of performance. All methods were applied using the
default values and an FDR cutoff of 0.01. Table~\ref{tabletotals}
indicates the number of significant peaks obtained in each case. The
methods are compared by a motif analysis and in terms of their mutual
agreement below.

\begin{table}[b]
\caption{Number of significant peaks called by all
methods at FDR level 0.01}\label{tabletotals}
\begin{tabular*}{\textwidth}{@{\extracolsep{\fill}}lcccc@{}}
\hline
\textbf{Dataset} & \textbf{STEM${}\bolds{+}{}$Regr} & \textbf{MACS} & \textbf{cisGenome} & \textbf{QuEST} \\
\hline
FoxA1 & 21,986 & 13,639 & 5725 & 20,161 \\
GABP & \phantom{.0}3309 & 13,828 & 4275 & \phantom{0.}6442\\
\hline
\end{tabular*}
\end{table}

\subsection{Motif analysis}

As biological validation, a motif analysis was performed where, for
each peak declared significant, the number of motifs related to the
appropriate transcription factor was counted within 100~bp and 400~bp
of the estimated peak location. The distance of 400~bp approximately
corresponds to the spatial spread of the measurements belonging to a
binding site, as determined by the estimate in Figure~\ref{figpeak-shape}(a).

Table~\ref{tablemotif-analysis} shows the average number of motifs and
the proportion of peaks with at least one motif within those distances
for the top 5725 peaks found by each method in the FoxA1 data set and
the top 3309 peaks found by each method in the GABP data set. These
numbers are the minima of the rows in Table~\ref{tabletotals}. Taking
the same number of top peaks in each list makes the averages and
proportions in the table comparable, as the peak lists are ordered and
the various methods use different criteria for their list cutoffs. Our
method, labeled ``STEM${}+{}$Regr'' for simplicity, shows a similar
performance to the other methods. Given the standard errors, it is
difficult to claim superiority of any method over the others.

\begin{table}
\caption{Motif analysis comparing the
performance of the proposed method against MACS and cisGenome on two
different data sets. Results are for the top 5725 peaks in each method
for the FoxA1 data set and the top 3309 peaks in each method for the
GABP data set. Standard errors are all between 1\% and 2\% of the
number shown}\label{tablemotif-analysis}
\begin{tabular*}{\textwidth}{@{\extracolsep{4in minus 4in}}lccccc@{}}
\hline
& & \multicolumn{2}{c}{\textbf{Average number of}} & \multicolumn
{2}{c@{}}{\textbf{Proportion with at}} \\
& & \multicolumn{2}{c}{\textbf{motifs within}} & \multicolumn{2}{c@{}}{\textbf{least one
motif within}} \\[-4pt]
& & \multicolumn{2}{c}{\hrulefill} & \multicolumn{2}{c@{}}{\hrulefill} \\
\textbf{Dataset} & \textbf{Method} & \textbf{100~bp} & \textbf{400~bp} & \textbf{100~bp} & \multicolumn{1}{c@{}}{\textbf{400~bp}} \\
\hline
FoxA1 & STEM${}+{}$Regr & 0.916 & 1.868 & 0.623 & 0.837 \\
& MACS & 0.917 & 1.849 & 0.625 & 0.835 \\
& cisGenome & 0.915 & 1.833 & 0.619 & 0.830 \\
& QuEST & 0.844 & 1.784 & 0.576 & 0.816 \\[3pt]
GABP & STEM${}+{}$Regr & 0.880 & 1.708 & 0.573 & 0.788 \\
& MACS & 0.875 & 1.703 & 0.579 & 0.792 \\
& cisGenome & 0.862 & 1.658 & 0.562 & 0.766 \\
& QuEST & 0.868 & 1.725 & 0.578 & 0.804\\
\hline
\end{tabular*}
\end{table}

\subsection{Peak overlap and discrepancies}

To help explain the previous results, Table~\ref{tableoverlap}
compares the percentage of peaks from the top 5725 from each method in
the FoxA1 data set or the top 3309 from each method in the GABP data
set, that were also found by each of the other methods within a
distance of 100~bp and 400~bp. The matrices in the table are not
symmetric because the correspondence between peaks is not one-to-one;
peaks found by one method may be represented by two or more peaks found
by another method. The table shows that there is a fair amount of
overlap between the methods, particularly in the GABP data set where
the peak lists are smaller (Table~\ref{tabletotals}).


\begin{table}
\tabcolsep=3pt
\caption{Percentage of peaks from the methods
listed in the columns that were also found by~the~methods listed in the
rows within a distance of 100~bp and 400~bp.\break Results are for the top
5725 peaks from each method in the FoxA1 data set\break and the top 3309
peaks from each method in the GABP data set}\label{tableoverlap}
\begin{tabular*}{\textwidth}{@{\extracolsep{\fill}}lcd{3.1}d{3.1}d{3.1}d{3.1}d{3.1}d{3.1}d{3.1}d{3.1}@{}}
\hline
& & \multicolumn{4}{c}{\textbf{\% found within 100~bp}} & \multicolumn{4}{c@{}}{\textbf{\% found within 400~bp}}
\\[-4pt]
& & \multicolumn{4}{c}{\hrulefill} & \multicolumn{4}{c@{}}{\hrulefill} \\
\textbf{Dataset} & \textbf{Method} & \multicolumn{1}{c}{\rotatebox{-90}{\textbf{STEM${}\bolds{+}{}$Regr}}} & \multicolumn{1}{c}{\rotatebox{-90}{\textbf{MACS}}} &
\multicolumn{1}{c}{\rotatebox{-90}{\textbf{cisGenome}}} &
\multicolumn{1}{c}{\rotatebox{-90}{\textbf{QuEST}}} & \multicolumn{1}{c}{\rotatebox{-90}{\textbf{STEM${}\bolds{+}{}$Regr}}} & \multicolumn{1}{c}{\rotatebox
{-90}{\textbf{MACS}}} & \multicolumn{1}{c}{\rotatebox{-90}{\textbf{cisGenome}}} & \multicolumn{1}{c}{\rotatebox{-90}{\textbf{QuEST}}} \\
\hline
FoxA1 & STEM${}+{}$Regr & 100 & 81.0 & 76.8 & 64.1 & 100 & 84.4 & 79.5 & 64.5
\\
& MACS & 79.6 & 100 & 84.4 & 70.8 & 80.1 & 100 & 84.9 & 71.0 \\
& cisGenome & 75.1 & 84.4 & 100 & 68.8 & 75.4 & 84.9 & 100 & 68.8 \\
& QuEST & 64.2 & 71.5 & 70.5 & 100 & 69.3 & 80.5 & 77.9 & 100 \\[3pt]
GABP & STEM${}+{}$Regr & 100 & 90.1 & 83.7 & 89.8 & 100 & 92.8 & 86.5 & 89.9
\\
& MACS & 90.1 & 100 & 84.7 & 87.8 & 90.4 & 100 & 85.8 & 87.9 \\
& cisGenome & 83.7 & 84.7 & 100 & 80.8 & 84.1 & 85.8 & 100 & 80.8 \\
& QuEST & 90.1 & 87.8 & 80.8 & 100 & 93.9 & 94.1 & 86.7 & 100\\
\hline
\end{tabular*}\vspace*{-3pt}
\end{table}

\begin{figure}

\includegraphics{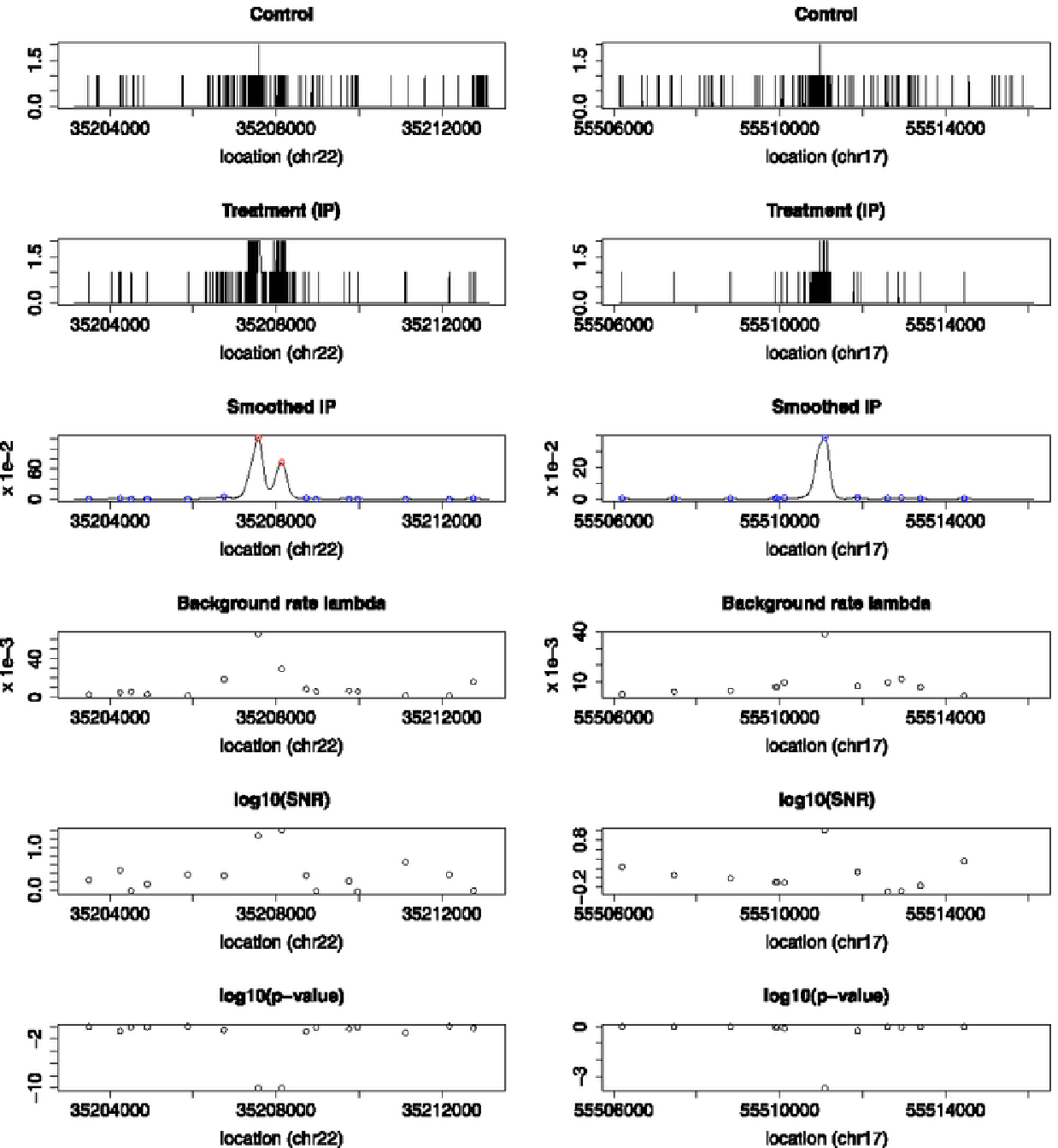}

\caption{Left: A fragment of the aligned GABP data featuring a
secondary peak called
by our method but not MACS or cisGenome. Right: A fragment of the
aligned GABP data featuring
a peak called by MACS and cisGenome but not by our method. The
variables plotted are the same as in
Figure \protect\ref{figFoxA1examples}. Notice the difference in
vertical scales between the left and right panels.}
\label{figGABPexamples}
\end{figure}

To better understand the discrepancies, Figure~\ref{figGABPexamples}
shows two examples of genomic segments from the GABP data set after
alignment. The left panel shows one of the 32 peaks produced by our
method that were not found among the peaks produced by MACS or
cisGenome. Our method detected a secondary peak within 567~bp of a
major peak (Row 3), in a binding region that was counted as a single
region by both MACS and cisGenome. All the other peaks in this group of
32 were found to be secondary peaks or sometimes tertiary peaks, with
distances between 385~bp and 1113~bp from their closest neighbor.

These secondary peaks, not distinguished by MACS or cisGenome, may be
separate binding sites. The ability to resolve them is a consequence of
our method searching for binding sites rather than binding regions.
These secondary sites were also found by QuEST, but were often
represented by perhaps too many peaks. For example, the secondary peak
in the left panel of Figure~\ref{figGABPexamples} was identified by
QuEST as two peaks, but being within only 131~bp of each other,
they may not belong to separate sites. The ability to represent a
single site by a single peak is a consequence of our method using a
matched filter rather than a narrow Gaussian filter.

The right panel of Figure~\ref{figGABPexamples} shows one of the 575
of the peaks that were produced by MACS and called by cis Genome and
QuEST but were not among the top 3309 produced by our method. This peak
was not called significant by our method because its associated $p$-value
was not low enough (Row 6). This is because the peak height is low (Row
3), while the estimated local background rate is high (Row 4),
resulting in a relatively low SNR (Row 5). Other peaks in this group of
575 were similar. This example illustrates the importance of the
estimation of the local background rate in the analysis.\looseness=1

\section{Simulations}
\label{secsimulations}

In order to evaluate the accuracy of the background rate estimation
method and the performance of the STEM algorithm for peak detection, we
performed the following spike-in simulated experiment. In each
simulated data set, two independent Poisson sequences of length
$L=10^7$ base pairs representing an aligned IP sequence and an aligned
control sequence were generated according to model \eqref
{eqPoisson-model}. The control background rate $\lambda_{\C}(t)$ was
obtained from chromosome~2 of the FoxA1 data set in a way similar to
model \eqref{eqlambda0} as
\[
\lambda_{\C}(t) = a_1 \hat{\lambda}_{\C,1k}(t) +
a_2 \hat{\lambda}_{\C,10k}(t),
\]
where $\hat{\lambda}_{\C,1k}(t)$ and $\hat{\lambda}_{\C,10k}(t)$ are
the Control averages in windows of size 1~Kb and 10~Kb centered at $t$,
$a_1 = 0.3$ and $a_2 = 0.7$. To simulate a different enrichment between
the IP and control sequences, the IP background rate was set to $\lambda_0(t) = 0.8 \lambda_C(t)$.
Then the actual IP rate was set to $\lambda_{IP}(t) = \lambda_0(t) + S \lambda^+(t)$, where $\lambda^+(t)$ is a
sequence of 20 spikes with shape equal to the solid curve in Figure \ref
{figpeak-shape} but normalized so that the area under each peak is
equal to the mean of $\lambda_0(t)$. Because of this normalization, the
factor $S$ can be interpreted as the signal-to-noise ratio and it was
set to values between 5 and 15.

\begin{figure}

\includegraphics{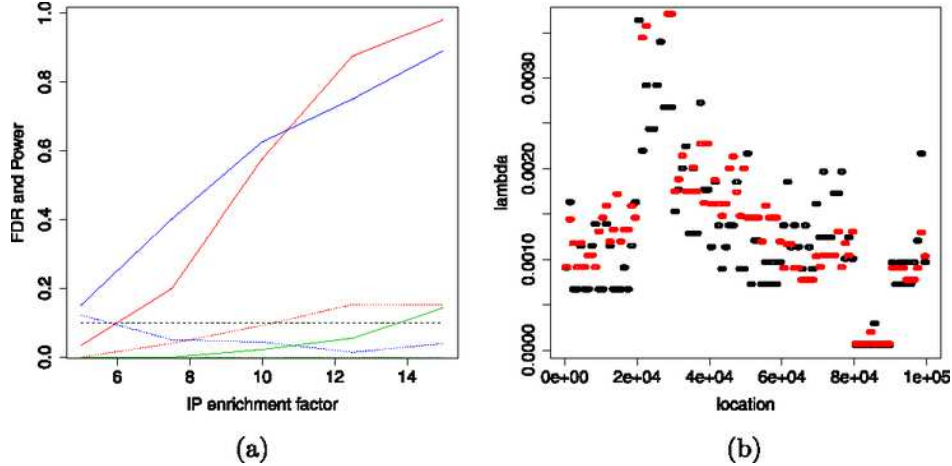}

\caption{Spike-in simulated experiment in a section of Chromosome 2.
\textup{(a)} Detection performance of the three methods: STEM${}+{}$Regr (red),
cisGenome (blue) and MACS (green).
Detection power is shown with solid lines, FDR with dotted lines.
Black dashed line is the nominal FDR level 0.1. \textup{(b)} Estimation of the
background rate $\lambda_0(t)$ as
a function of genomic location $t$ (selected fragment): simulated
(black) and estimated (red).}
\label{figsim}
\end{figure}

Figure~\ref{figsim}(a) shows the realized FDR and detection power
(defined as the fraction of detected peaks) averaged over 10
independent data sets simulated as described above. The proposed
STEM${}+{}$Regr algorithm shows similar performance and error control as
cisGenome. MACS's apparent low power may be rather an indication that
the algorithm is not intended to be applied to short sequences like the
ones used in this simulated experiment. Results for QuEST were not
obtained because the need for user input makes the software not
conducive for repeated simulation experiments of this kind.

Looking closer at the regression method for estimating the background
rate $\lambda_0(t)$, Figure~\ref{figsim}(b) shows that the regression
method is able to follow the general trend of the local background rate
despite it varying quickly. To reduce variability, the method
automatically performs a bias-variance trade-off, where the
coefficients $C \times a_1 = 0.24$ and $C \times a_2 = 0.56$ are
estimated on average as 0.096 and 0.678, respectively. The overall
correlation between the simulated and predicted background rate is 0.73.\

\section{Discussion}
\label{secdiscussion}

\subsection{Methodological considerations}
We have presented a method for detection of peaks in ChIP-Seq data
based on the STEM algorithm of SGA with promising results. The
applicability of SGA to ChIP-Seq data relied on the common assumption
that the signal peaks, represented by a mean function, are unimodal and
have the same shape up to an amplitude scaling factor. The adaptation
to ChIP-Seq data required two main modifications: (1) estimation of the
local background rate; (2) use of Monte Carlo simulations of Poisson
sequences to compute $p$-values.

From a methodological point of view, estimation of the background rate
$\lambda_0(t)$ is arguably the most crucial step in the analysis, as
the inference for a particular local maximum is highly dependent on the
background rate at that location. In this paper, we have focused on the
inference aspects of detecting peaks with a spatial structure via the
STEM algorithm. Estimation of the local background rate (the particle
``Regr'' in the acronym ``STEM${}+{}$Regr'') is not part of the original STEM
algorithm, but is necessary for the analysis of ChIP-Seq data because
the noise process is not stationary. This conceptual separation is
helpful in that the background estimation method could be replaced by a
different method if desired, without affecting the general
implementation of the STEM algorithm for peak detection.

Statistical methods for estimating the variable rate $\lambda(t)$ in
dynamic Poisson models or nonhomogeneous Poisson sequences have been
developed in other contexts. Bayesian methods [\citet
{West1985,Harvey1986,Bolstad1995}] are computationally intensive,
estimating $\lambda(t)$ at each location $t$ based on the estimates at
locations $1,2,\ldots,t-1$. This is computationally infeasible for long
genomic sequences as in ChIP-Seq data. Other methods require either
repeated realizations [\citet{Arkin2000}] or a more specific structure
of the process [\citet{Zhao1996,Helmers2003}], which are not available
in ChIP-Seq data.

In this paper we have proposed a simple solution to local background
estimation based on multiple linear regression using the Control sample
as a covariate. If desired, other covariates could be included such as
other window sizes and the local GC content. Estimating the regression
coefficients from the data automatically adjusts for sequencing depth,
and the estimated relative weighting between the various window sizes
allows the method to adaptively estimate\vadjust{\goodbreak} the local background at each
location. In this sense, the regression model solves the normalization
problem and gives a partial answer to the question of how slowly
$\lambda_0(t)$ varies with $t$. Because the GABP data set is richer in
number of reads, the regression method automatically accounts for it
and allows estimation of the background rate at a smaller spatial scale
by giving a higher weight to the 1~kb window relative to the 10~kb
window than in the FoxA1 data set.

Often in ChIP-Seq data a Control sample is unavailable. In such cases,
the regression model \eqref{eqlambda0} could have the 10~Kb averages
from the IP sample itself as predictors instead of those from the
Control, with the 1~Kb window not included in the model. This would
allow estimation of the background from the neighborhood of each peak,
albeit with some positive bias. Fortunately, the positive bias would
make the inference more conservative, affecting the detection power
more than its validity.

In the comparison with the other methods, it was observed that the STEM
algorithm performs competitively in terms of nearby motifs. In the data
sets analyzed, all methods found many of the same strong peaks.
However, our method found secondary and terciary peaks near other
strong peaks that were not distinguished by MACS and cisGenome and were
too fragmented by QuEST. This is a result of our method searching for
localized binding sites using an appropriately chosen matched filter
rather than binding regions of arbitrary size.

On the other hand, our method did not call significant other peaks that
were called by the other methods. These peaks were not strong enough
when compared to their corresponding background estimate at that
location, at least according to the background estimation method used
here. It is possible that a different background estimation method
would have caused these peaks to be called significant. In fact, the
other methods did because they had different assumptions about what
represents a strong peak.

In this paper, we have attempted to frame the ChIP-Seq problem as a
formal multiple testing problem. The significance results and FDR
levels may be trusted under the proposed model, however, the biological
validity of the results is limited by the validity of the modeling
assumptions. Particularly difficult is the background estimation, for
which no good model exists to date. Because of its importance,
background estimation is where future research in ChIP-Seq analysis
should focus its attention.

\subsection{Computational considerations}

In addition to the modelling considerations mentioned above, the final
ranking of the detected peaks depends on the numerical accuracy with
which $p$-values are computed. In the Monte Carlo simulations for
computing the distribution of the heights of local maxima, the
estimation is more accurate for high values of $\lambda$, as these
produce more observations. In the simulations, we set the simulation
length to be at least $10^5$, or as long as is needed to obtain at
least 100 Poisson counts. The latter\vadjust{\goodbreak} condition was necessary for very
low Poisson rates, but cannot be considered sufficient. The random
variability was attenuated by B-spline smoothing across $\lambda$ in
order to obtain the table in Figure~\ref{figlambda-distr}(b).

In the analysis results, we observed that a large number of detected
peaks had a $p$-value of zero, meaning that the Monte Carlo simulation
did not have enough numerical resolution to distinguish between their
$p$-values. These peaks were ranked sub-optimally by SNR. More accurate
calculation of $p$-values could be achieved with longer Monte Carlo
simulations or by more sophisticated simulation techniques, such as
Importance Sampling.

Computational complexity is also important in ChIP-Seq analysis because
of the large amount of data to be processed. The methods in this paper
were implemented in \texttt{R} to ease their development and sharing among
researchers, but at the expense of computational speed. The main
computational bottleneck of our method is kernel smoothing, taking
about $6\sim8$ hours to run over the entire genome on a Dell Power
Edge R710 server with CPU speed 2.67 GHz, 48 GB of memory and a Linux
CentOS 5.5 operating system. All the other processing steps together
take about another hour. Kernel smoothing is mathematically simple, yet
unfortunately inefficient in \texttt{R} for very long sequences. Computing
time for kernel smoothing increases linearly with the kernel and the
sequence size. In our implementation, the data was divided into
subgroups of tags no more than $10^4$~bp apart, trading off the length
of the groups and their number. Computational time was also reduced by
reducing the length of the kernel by multiplying it by a quartic
biweight function of smaller support and using run length encoding in
the search for local maxima. In the future the ideas proposed here
could be made computationally competitive by implementing them in \texttt{C}.

We do not intend that the method proposed in this paper is viewed as a
competitor to other existing methods for analyzing ChIP-Seq data, but
rather as a suggestion of how multiple testing theory for spatial
domains, such as in SGA, can inform the inference procedure in the
detection of peaks. While competitive in terms of detection
performance, the strength of our method relies mainly on the potential
generalization of these ideas to other domains in spatial inference,
both in bioinformatics and beyond.


\begin{appendix}\label{app}
\section*{Appendix: Alignment details}
\label{secdata}

\subsection{Raw data}
The FoxA1 raw data consists of a table of about 3.9 million rows for
the IP sample and a table of about 5.2 million rows for the Control
sample. Each row corresponds to a mapped tag of length 35~bp and
contains the beginning and end genomic addresses for the tag and an
indicator of whether the tag belongs to the forward ($+$) or reverse
($-$) DNA strand. We define the location of a tag to be given by its
beginning address, corresponding to the lower address for the forward
($+$) tags and the higher address for the reverse ($-$) tags. The GABP
data set, containing about 7.8 million\vadjust{\goodbreak} tags in the IP sample and about
17.4 million tags in the Control sample, was converted to the same
format before processing. Genomic locations not listed in the table
were assumed to have an associated tag count of zero. Duplicate tags
were considered measurement artifacts and were removed from the analysis.

In order to be counted together, the tags from the two strands need to
be aligned with each other. We followed an alignment method similar to
that in MACS, shifting all tags by the same amount in the 3' direction
of the tag sequence toward the most likely binding site: forward $(+)$
tags toward higher genomic addresses and reverse $(-)$ tags toward
lower genomic addresses. Once shifted, tags coinciding at the same
location are counted together. The result of this process is a table of
genomic locations, each with an associated tag count. This aligned data
is used as the input for peak detection, described in Section~\ref{secmain}.

\subsection{Estimation of the tag shift and peak shape}

As in MACS, we estimate the size of the shift from the tag count
distributions corresponding to a set of strong and easily detectable
peaks, as described below. We performed the shift estimation on
Chromosome 1 because of its likelihood to contain enough such strong
peaks, but other long chromosomes could be used instead. As part of the
process, the shift estimation also allows us to estimate the
distribution of shifted tags counts around a peak. This peak shape,
normalized to unit sum, is used later as a smoothing kernel in the STEM
algorithm for peak detection. The estimation proceeds as follows.

\begin{alg}[(Estimation of shift size and peak shape)]
\label{algdistr}
\begin{longlist}[(1)]
\item[(1)] Temporarily shift all tags [$(+)$ forward and $(-)$ back] by a
tentative shift amount (default 100~bp). This produces a table of
genomic locations, each with an associated tag count.
\item[(2)] Perform peak detection on the count data from the previous step
and select a set of strong peaks (details given below). Let $t_1,\ldots
,t_N$ be their locations.
\item[(3)] Set a window size $W$ (an odd number, default 2001~bp). The
distribution of the forward tags is a vector of length $W$ whose $i$th
entry is equal to the average number of forward tags at a constant
distance $(W+1)/2 - i$ from the peak, that is, at locations $t_j -
(W+1)/2 + i$, $j=1,\ldots,N$. Repeat for the reverse tags.
\item[(4)] Fit a spline to the distribution of forward tags and record its
mode. Repeat for the reverse tags. The estimated shift is half the
distance between the two modes, rounded to the nearest integer.
\item[(5)] To estimate the peak shape, shift the original forward and
reverse distributions by the estimated shift, symmetrize the joint
distribution by averaging both the forward and reverse tag
distributions and their mirror images with respect to the center of the
window, and fit a spline.
\end{longlist}
\end{alg}

The peak detection step (Step 2) above need not be exact. Since the tag
distribution is evaluated in a window around the strong peaks, it is
enough that the true location of those peaks is contained somewhere
near the center of that window. To achieve this, we apply the first
half of the STEM algorithm, as follows.
\begin{longlist}[(2a)]
\item[(2a)] Set a tentative unimodal symmetric kernel (default Gaussian
with standard deviation 50) and perform kernel smoothing on the count
data from Step 1. (Implementation details given in Section~\ref{secsmoothing}).
\item[(2b)] Find the local maxima of the smoothed count sequence.
(Implementation details given in Section~\ref{secsmoothing}).
\item[(2c)] Select the $N$ highest local maxima (default 1000).
\end{longlist}

At the end of this process, the data consists of a long sequence of
genomic addresses and associated counts 0, 1 or 2, ready for peak
detection analysis. The maximal count of 2 is a result of the
elimination of duplicates from the original list of tags. Because
binding rates are generally low, truncation at 2 does not greatly
affect the Poisson model used thereafter.
\end{appendix}

%


\printaddresses


\begin{thebibliography}{27}

\bibitem[\protect\citeauthoryear{Arkin and Leenis}{2000}]{Arkin2000}
%
\begin{barticle}[author]
\bauthor{\bsnm{Arkin},~\bfnm{Bradford~L.}\binits{B.~L.}} \AND
\bauthor{\bsnm{Leenis},~\bfnm{Lawrence~M.}\binits{L.~M.}}
(\byear{2000}).
\btitle{Nonparametric estimation of the cumulative intensity function
for a
nonhomogeneous {P}oisson process from overlapping realizations}.
\bjournal{Management Science}
\bvolume{46}
\bpages{989--998}.
\bptok{imsref}%
\end{barticle}
%
\endbibitem

\bibitem[\protect\citeauthoryear{Barski and Zhao}{2009}]{Barski2009}
%
\begin{barticle}[pbm]
\bauthor{\bsnm{Barski},~\bfnm{Artem}\binits{A.}} \AND
\bauthor{\bsnm{Zhao},~\bfnm{Keji}\binits{K.}}
(\byear{2009}).
\btitle{Genomic location analysis by ChIP-Seq}.
\bjournal{J. Cell. Biochem.}
\bvolume{107}
\bpages{11--18}.
\bid{doi={10.1002/jcb.22077}, issn={1097-4644}, pmid={19173299}}
\bptok{imsref}%
\end{barticle}
%
\endbibitem

\bibitem[\protect\citeauthoryear{Benjamini and Hochberg}{1995}]{Benjamini1995}
%
\begin{barticle}[mr]
\bauthor{\bsnm{Benjamini},~\bfnm{Yoav}\binits{Y.}} \AND
\bauthor{\bsnm{Hochberg},~\bfnm{Yosef}\binits{Y.}}
(\byear{1995}).
\btitle{Controlling the false discovery rate: A practical and powerful approach
to multiple testing}.
\bjournal{J. Roy. Statist. Soc. Ser. B}
\bvolume{57}
\bpages{289--300}.
\bid{issn={0035-9246}, mr={1325392}}
\bptok{imsref}%
\end{barticle}
%
\endbibitem

\bibitem[\protect\citeauthoryear{Bolstad}{1995}]{Bolstad1995}
%
\begin{barticle}[mr]
\bauthor{\bsnm{Bolstad},~\bfnm{William~M.}\binits{W.~M.}}
(\byear{1995}).
\btitle{The multiprocess dynamic {P}oisson model}.
\bjournal{J. Amer. Statist. Assoc.}
\bvolume{90}
\bpages{227--232}.
\bid{issn={0162-1459}, mr={1325130}}
\bptok{imsref}%
\end{barticle}
%
\endbibitem

\bibitem[\protect\citeauthoryear{Fejes et~al.}{2008}]{Fejes2008}
%
\begin{barticle}[author]
\bauthor{\bsnm{Fejes},~\bfnm{A.}\binits{A.}},
\bauthor{\bsnm{Robertson},~\bfnm{G.}\binits{G.}},
\bauthor{\bsnm{Bilenky},~\bfnm{M.}\binits{M.}},
\bauthor{\bsnm{Varhol},~\bfnm{R.}\binits{R.}},
\bauthor{\bsnm{Bainbridge},~\bfnm{M.}\binits{M.}} \AND
\bauthor{\bsnm{Jones},~\bfnm{S.}\binits{S.}}
(\byear{2008}).
\btitle{{F}ind{P}eaks 3.1: A tool for identifying areas of enrichment from
massively parallel short-read sequencing technology}.
\bjournal{Bioinformatics}
\bvolume{24}
\bpages{1720--1730}.
\bptok{imsref}%
\end{barticle}
%
\endbibitem

\bibitem[\protect\citeauthoryear{Harvey and Durbin}{1986}]{Harvey1986}
%
\begin{barticle}[author]
\bauthor{\bsnm{Harvey},~\bfnm{A.~C.}\binits{A.~C.}} \AND
\bauthor{\bsnm{Durbin},~\bfnm{J.}\binits{J.}}
(\byear{1986}).
\btitle{The effects of seat belt legislation on {B}ritish road
casualties: A
case study in structural time series modelling}.
\bjournal{J. Roy. Statist. Soc.}
\bvolume{149}
\bpages{187--227}.
\bptok{imsref}%
\end{barticle}
%
\endbibitem

\bibitem[\protect\citeauthoryear{Helmers, Mangku and
Zitikis}{2003}]{Helmers2003}
%
\begin{barticle}[mr]
\bauthor{\bsnm{Helmers},~\bfnm{Roelof}\binits{R.}},
\bauthor{\bsnm{Mangku},~\bfnm{I.~Wayan}\binits{I.~W.}} \AND
\bauthor{\bsnm{Zitikis},~\bfnm{Ri{\v{c}}ardas}\binits{R.}}
(\byear{2003}).
\btitle{Consistent estimation of the intensity function of a cyclic {P}oisson
process}.
\bjournal{J. Multivariate Anal.}
\bvolume{84}
\bpages{19--39}.
\bid{doi={10.1016/S0047-259X(02)00008-8}, issn={0047-259X}, mr={1965821}}
\bptok{imsref}%
\end{barticle}
%
\endbibitem

\bibitem[\protect\citeauthoryear{Hower, Evans and Pachter}{2011}]{T-PIC2011}
%
\begin{barticle}[author]
\bauthor{\bsnm{Hower},~\bfnm{Valerie}\binits{V.}},
\bauthor{\bsnm{Evans},~\bfnm{Steve~N.}\binits{S.~N.}} \AND
\bauthor{\bsnm{Pachter},~\bfnm{Lior}\binits{L.}}
(\byear{2011}).
\btitle{Shape-based identification for {ChIP-Seq}}.
\bjournal{BMC Bioinformatics}
\bvolume{12}
\bpages{15}.
\bptok{imsref}%
\end{barticle}
%
\endbibitem

\bibitem[\protect\citeauthoryear{Jaffe et~al.}{2012}]{Jaffe2012}
%
\begin{barticle}[pbm]
\bauthor{\bsnm{Jaffe},~\bfnm{Andrew~E.}\binits{A.~E.}},
\bauthor{\bsnm{Feinberg},~\bfnm{Andrew~P.}\binits{A.~P.}},
\bauthor{\bsnm{Irizarry},~\bfnm{Rafael~A.}\binits{R.~A.}} \AND
\bauthor{\bsnm{Leek},~\bfnm{Jeffrey~T.}\binits{J.~T.}}
(\byear{2012}).
\btitle{Significance analysis and statistical dissection of variably methylated
regions}.
\bjournal{Biostatistics}
\bvolume{13}
\bpages{166--178}.
\bid{doi={10.1093/biostatistics/kxr013}, issn={1468-4357}, pii={kxr013},
pmcid={3276267}, pmid={21685414}}
\bptok{imsref}%
\end{barticle}
%
\endbibitem

\bibitem[\protect\citeauthoryear{Ji et~al.}{2008}]{cisGenome2008}
%
\begin{barticle}[author]
\bauthor{\bsnm{Ji},~\bfnm{Hongkai}\binits{H.}},
\bauthor{\bsnm{Jiang},~\bfnm{Hui}\binits{H.}},
\bauthor{\bsnm{Ma},~\bfnm{Wenxiu}\binits{W.}},
\bauthor{\bsnm{Johnson},~\bfnm{David~S.}\binits{D.~S.}},
\bauthor{\bsnm{Myers},~\bfnm{Richard~M.}\binits{R.~M.}} \AND
\bauthor{\bsnm{Wong},~\bfnm{Wing~H.}\binits{W.~H.}}
(\byear{2008}).
\btitle{An integrated software system for analyzing {ChIP-chip} and {ChIP-Seq}
data}.
\bjournal{Nature Biotechnology}
\bvolume{26}
\bpages{1293--1300}.
\bptok{imsref}%
\end{barticle}
%
\endbibitem

\bibitem[\protect\citeauthoryear{Johnson et~al.}{2007}]{Johnson2007}
%
\begin{barticle}[pbm]
\bauthor{\bsnm{Johnson},~\bfnm{David~S.}\binits{D.~S.}},
\bauthor{\bsnm{Mortazavi},~\bfnm{Ali}\binits{A.}},
\bauthor{\bsnm{Myers},~\bfnm{Richard~M.}\binits{R.~M.}} \AND
\bauthor{\bsnm{Wold},~\bfnm{Barbara}\binits{B.}}
(\byear{2007}).
\btitle{Genome-wide mapping of in vivo protein-DNA interactions}.
\bjournal{Science}
\bvolume{316}
\bpages{1497--1502}.
\bid{doi={10.1126/science.1141319}, issn={1095-9203}, pii={1141319},
pmid={17540862}}
\bptok{imsref}%
\end{barticle}
%
\endbibitem

\bibitem[\protect\citeauthoryear{Mikkelsen et~al.}{2007}]{Mikkelsen2007}
%
\begin{barticle}[author]
\bauthor{\bsnm{Mikkelsen},~\bfnm{T.~S.}\binits{T.~S.}},
\bauthor{\bsnm{Ku},~\bfnm{M.}\binits{M.}},
\bauthor{\bsnm{Jaffe},~\bfnm{D.~B.}\binits{D.~B.}},
\bauthor{\bsnm{Issac},~\bfnm{B.}\binits{B.}},
\bauthor{\bsnm{Lieberman},~\bfnm{E.}\binits{E.}},
\bauthor{\bsnm{Giannoukos},~\bfnm{G.}\binits{G.}},
\bauthor{\bsnm{Alvarez},~\bfnm{P.}\binits{P.}},
\bauthor{\bsnm{Brockman},~\bfnm{W.}\binits{W.}},
\bauthor{\bsnm{Kim},~\bfnm{T.~K.}\binits{T.~K.}},
\bauthor{\bsnm{Koche},~\bfnm{R.~P.}\binits{R.~P.}},
\bauthor{\bsnm{Lee},~\bfnm{W.}\binits{W.}},
\bauthor{\bsnm{Mendenhall},~\bfnm{E.}\binits{E.}},
\bauthor{\bsnm{O'Donovan},~\bfnm{A.}\binits{A.}},
\bauthor{\bsnm{Presser},~\bfnm{A.}\binits{A.}},
\bauthor{\bsnm{Russ},~\bfnm{C.}\binits{C.}},
\bauthor{\bsnm{Xie},~\bfnm{X.}\binits{X.}},
\bauthor{\bsnm{Meissner},~\bfnm{A.}\binits{A.}},
\bauthor{\bsnm{Wernig},~\bfnm{M.}\binits{M.}},
\bauthor{\bsnm{Jaenisch},~\bfnm{R.}\binits{R.}},
\bauthor{\bsnm{Nusbaum},~\bfnm{C.}\binits{C.}},
\bauthor{\bsnm{Lander},~\bfnm{E.~S.}\binits{E.~S.}} \AND
\bauthor{\bsnm{Bernstein},~\bfnm{B.~E.}\binits{B.~E.}}
(\byear{2007}).
\btitle{Genome-wide maps of chromatin state in pluripotent and
lineage-committed cells}.
\bjournal{Nature}
\bvolume{448}
\bpages{553--560}.
\bptok{imsref}%
\end{barticle}
%
\endbibitem

\bibitem[\protect\citeauthoryear{North}{1943}]{North1943}
%
\begin{bmisc}[author]
\bauthor{\bsnm{North},~\bfnm{D.~O.}\binits{D.~O.}}
(\byear{1943}).
\bhowpublished{An analysis of the factors which determine signal/noise
discrimination
in pulsed carrier systems. Technical Report No. PTR-6C, RCA Labs,
Princeton, NJ.}
\bptok{imsref}%
\end{bmisc}
%
\endbibitem

\bibitem[\protect\citeauthoryear{Park}{2009}]{Park2009}
%
\begin{barticle}[pbm]
\bauthor{\bsnm{Park},~\bfnm{Peter~J.}\binits{P.~J.}}
(\byear{2009}).
\btitle{ChIP-seq: Advantages and challenges of a maturing technology}.
\bjournal{Nat. Rev. Genet.}
\bvolume{10}
\bpages{669--680}.
\bid{doi={10.1038/nrg2641}, issn={1471-0064}, mid={NIHMS327265}, pii={nrg2641},
pmcid={3191340}, pmid={19736561}}
\bptok{imsref}%
\end{barticle}
%
\endbibitem

\bibitem[\protect\citeauthoryear{Pratt}{1991}]{Pratt1991}
%
\begin{bbook}[author]
\bauthor{\bsnm{Pratt},~\bfnm{William~K.}\binits{W.~K.}}
(\byear{1991}).
\btitle{Digital Image Processing}.
\bpublisher{Wiley}, \blocation{New York}.
\bptok{imsref}%
\end{bbook}
%
\endbibitem

\bibitem[\protect\citeauthoryear{Sakharkar, Chow and
Kangueane}{2004}]{Sakharkar2004}
%
\begin{barticle}[author]
\bauthor{\bsnm{Sakharkar},~\bfnm{Meena~Kishore}\binits{M.~K.}},
\bauthor{\bsnm{Chow},~\bfnm{Vincent T.~K.}\binits{V.~T.~K.}} \AND
\bauthor{\bsnm{Kangueane},~\bfnm{Pandjassarame}\binits{P.}}
(\byear{2004}).
\btitle{Distributions of exons and introns in the human genome}.
\bjournal{In Silico Biology}
\bvolume{4}
\bpages{0032}.
\bptok{imsref}%
\end{barticle}
%
\endbibitem

\bibitem[\protect\citeauthoryear{Schwartzman, Gavrilov and
Adler}{2011}]{Schwartzman2011b}
%
\begin{barticle}[author]
\bauthor{\bsnm{Schwartzman},~\bfnm{Armin}\binits{A.}},
\bauthor{\bsnm{Gavrilov},~\bfnm{Yulia}\binits{Y.}} \AND
\bauthor{\bsnm{Adler},~\bfnm{Robert~J.}\binits{R.~J.}}
(\byear{2011}).
\btitle{Multiple testing of local maxima for detection of peaks in 1D}.
\bjournal{Ann. Statist.}
\bvolume{39}
\bpages{3290--3319}.
\bid{mr={3012409}}
\bptok{imsref}%
\end{barticle}
%
\endbibitem

\bibitem[\protect\citeauthoryear{Simon}{1995}]{Simon1995}
%
\begin{bbook}[author]
\bauthor{\bsnm{Simon},~\bfnm{Marvin}\binits{M.}}
(\byear{1995}).
\btitle{Digital Communication Techniques: Signal Design and Detection}.
\bpublisher{Prentice Hall}, \blocation{Englewood Cliffs, NJ}.
\bptok{imsref}%
\end{bbook}
%
\endbibitem

\bibitem[\protect\citeauthoryear{Spyrou et~al.}{2009}]{BayesPeak2009}
%
\begin{barticle}[author]
\bauthor{\bsnm{Spyrou},~\bfnm{C.}\binits{C.}},
\bauthor{\bsnm{Stark.},~\bfnm{R.}\binits{R.}},
\bauthor{\bsnm{Lynch},~\bfnm{A.~G.}\binits{A.~G.}} \AND
\bauthor{\bsnm{Tavar},~\bfnm{S.}\binits{S.}}
(\byear{2009}).
\btitle{{B}ayesian analysis of {ChIP-Seq} data}.
\bjournal{BMC Bioinformatics}
\bvolume{10}
\bpages{299}.
\bptok{imsref}%
\end{barticle}
%
\endbibitem

\bibitem[\protect\citeauthoryear{Turin}{1960}]{Turin1960}
%
\begin{barticle}[mr]
\bauthor{\bsnm{Turin},~\bfnm{George~L.}\binits{G.~L.}}
(\byear{1960}).
\btitle{An introduction to matched filters}.
\bjournal{Trans. IRE}
\bvolume{IT-6}
\bpages{311--329}.
\bid{mr={0115847}}
\bptok{imsref}%
\end{barticle}
%
\endbibitem

\bibitem[\protect\citeauthoryear{Valouev et~al.}{2008}]{QuEST2008}
%
\begin{barticle}[author]
\bauthor{\bsnm{Valouev},~\bfnm{A.}\binits{A.}},
\bauthor{\bsnm{Johnson},~\bfnm{D.~S.}\binits{D.~S.}},
\bauthor{\bsnm{Sundquist},~\bfnm{A.}\binits{A.}},
\bauthor{\bsnm{Medina},~\bfnm{C.}\binits{C.}},
\bauthor{\bsnm{Anton},~\bfnm{E.}\binits{E.}},
\bauthor{\bsnm{Batzglou},~\bfnm{S.}\binits{S.}},
\bauthor{\bsnm{Myers},~\bfnm{R.~M.}\binits{R.~M.}} \AND
\bauthor{\bsnm{Sidow},~\bfnm{A.}\binits{A.}}
(\byear{2008}).
\btitle{Genome-wide analysis of transcription factor binding sites
based on
{ChIP-Seq} data}.
\bjournal{Nature Methods}
\bvolume{5}
\bpages{829--834}.
\bptok{imsref}%
\end{barticle}
%
\endbibitem

\bibitem[\protect\citeauthoryear{Wasserman}{2006}]{Wasserman2006}
%
\begin{bbook}[mr]
\bauthor{\bsnm{Wasserman},~\bfnm{Larry}\binits{L.}}
(\byear{2006}).
\btitle{All of Nonparametric Statistics}.
\bpublisher{Springer}, \blocation{New York}.
\bid{mr={2172729}}
\bptok{imsref}%
\end{bbook}
%
\endbibitem

\bibitem[\protect\citeauthoryear{West, Harrison and Migon}{1985}]{West1985}
%
\begin{barticle}[mr]
\bauthor{\bsnm{West},~\bfnm{Mike}\binits{M.}},
\bauthor{\bsnm{Harrison},~\bfnm{P.~Jeff}\binits{P.~J.}} \AND
\bauthor{\bsnm{Migon},~\bfnm{H{\'e}lio~S.}\binits{H.~S.}}
(\byear{1985}).
\btitle{Dynamic generalized linear models and {B}ayesian forecasting}.
\bjournal{J. Amer. Statist. Assoc.}
\bvolume{80}
\bpages{73--97}.
\bid{issn={0162-1459}, mr={0786598}}
\bptok{imsref}%
\end{barticle}
%
\endbibitem

\bibitem[\protect\citeauthoryear{Zhang et~al.}{2008}]{MACS2008}
%
\begin{barticle}[author]
\bauthor{\bsnm{Zhang},~\bfnm{Yong}\binits{Y.}},
\bauthor{\bsnm{Liu},~\bfnm{Tao}\binits{T.}},
\bauthor{\bsnm{Meyer},~\bfnm{Clifford~A.}\binits{C.~A.}},
\bauthor{\bsnm{Eeckhoute},~\bfnm{J{\'{e}}r{\^{o}}me}\binits{J.}},
\bauthor{\bsnm{Johnson},~\bfnm{David~S.}\binits{D.~S.}},
\bauthor{\bsnm{Bernstein},~\bfnm{Bradley~E.}\binits{B.~E.}},
\bauthor{\bsnm{Nussbaum},~\bfnm{Chad}\binits{C.}},
\bauthor{\bsnm{Myers},~\bfnm{Richard~M.}\binits{R.~M.}},
\bauthor{\bsnm{Brown},~\bfnm{Myles}\binits{M.}},
\bauthor{\bsnm{Li},~\bfnm{Wei}\binits{W.}} \AND
\bauthor{\bsnm{Liu},~\bfnm{X.~Shirley}\binits{X.~S.}}
(\byear{2008}).
\btitle{Model-based analysis of {ChIP-Seq} ({MACS})}.
\bjournal{Genome Biology}
\bvolume{9}
\bpages{R137}.
\bptok{imsref}%
\end{barticle}
%
\endbibitem

\bibitem[\protect\citeauthoryear{Zhang et~al.}{2011}]{PICS2010}
%
\begin{barticle}[mr]
\bauthor{\bsnm{Zhang},~\bfnm{Xuekui}\binits{X.}},
\bauthor{\bsnm{Robertson},~\bfnm{Gordon}\binits{G.}},
\bauthor{\bsnm{Krzywinski},~\bfnm{Martin}\binits{M.}},
\bauthor{\bsnm{Ning},~\bfnm{Kaida}\binits{K.}},
\bauthor{\bsnm{Droit},~\bfnm{Arnaud}\binits{A.}},
\bauthor{\bsnm{Jones},~\bfnm{Steven}\binits{S.}} \AND
\bauthor{\bsnm{Gottardo},~\bfnm{Raphael}\binits{R.}}
(\byear{2011}).
\btitle{P{ICS}: Probabilistic inference for {C}h{IP}-seq}.
\bjournal{Biometrics}
\bvolume{67}
\bpages{151--163}.
\bid{doi={10.1111/j.1541-0420.2010.01441.x}, issn={0006-341X}, mr={2898827}}
\bptnote{check year}%
\bptok{imsref}%
\end{barticle}
%
\endbibitem

\bibitem[\protect\citeauthoryear{Zhao and Xie}{1996}]{Zhao1996}
%
\begin{barticle}[mr]
\bauthor{\bsnm{Zhao},~\bfnm{M.}\binits{M.}} \AND
\bauthor{\bsnm{Xie},~\bfnm{M.}\binits{M.}}
(\byear{1996}).
\btitle{On maximum likelihood estimation for a general non-homogeneous
{P}oisson process}.
\bjournal{Scand. J. Stat.}
\bvolume{23}
\bpages{597--607}.
\bid{issn={0303-6898}, mr={1439714}}
\bptok{imsref}%
\end{barticle}
%
\endbibitem

\end{thebibliography}
\end{document}